\documentclass{ws-ijmpe}
\usepackage[super,compress]{cite}
\begin{document}

\markboth{Bojana Ilic, Dusan Zigic, Marko Djordjevic, Magdalena Djordjevic}{Utilizing high-$p_\perp$ theory and data to constrain the initial stages of QGP}

\catchline{}{}{}{}{}

\title{UTILIZING HIGH-$p_\perp$ THEORY AND DATA TO CONSTRAIN THE INITIAL STAGES OF QUARK-GLUON PLASMA}

\author{BOJANA ILIC}

\address{Institute of Physics Belgrade, University of Belgrade, Pregrevica 118\\
Belgrade, 11080,
Serbia\\
bojanab@ipb.ac.rs}

\author{DUSAN ZIGIC}

\address{Institute of Physics Belgrade, University of Belgrade, Pregrevica 118\\
Belgrade, 11080,
Serbia\\
zigic@ipb.ac.rs}

\author{MARKO DJORDJEVIC}

\address{Faculty of Biology, University of Belgrade, Studentski trg 16\\
Belgrade, 11000,
Serbia\\
dmarko@bio.bg.ac.rs}

\author{MAGDALENA DJORDJEVIC}

\address{Institute of Physics Belgrade, University of Belgrade, Pregrevica 118\\
Belgrade, 11080,
Serbia\\
magda@ipb.ac.rs}

\maketitle

\begin{history}
\received{Day Month Year}
\revised{Day Month Year}
\end{history}

\begin{abstract}
The scarce knowledge of the initial stages of quark-gluon plasma before the thermalization is mostly inferred through the low-$p_\perp$ sector. We propose a complementary approach in this report - the use of high-$p_\perp$ probes' energy loss. We study the effects of four commonly assumed initial stages, whose temperature profiles differ only before the thermalization, on high-$p_\perp$ $R_{AA}$ and $v_2$ predictions. The predictions are based on our Dynamical Radiative and Elastic ENergy-loss Approach (DREENA)  framework. We report insensitivity of $v_2$ to the initial stages, making it unable to distinguish between different cases. $R_{AA}$ displays sensitivity to the presumed initial stages, but current experimental precision does not allow resolution between these cases. We further revise the commonly accepted procedure of fitting the energy loss parameters, for each individual initial stage, to the measured $R_{AA}$. We show that the sensitivity of $v_2$ to various initial stages obtained through such procedure is mostly a consequence of fitting procedure, which may obscure the physical interpretations. Overall, the simultaneous study of high-$p_\perp$ observables, with unchanged energy loss parametrization and restrained temperature profiles, is crucial for future constraints on initial stages.
\end{abstract}


\ccode{PACS numbers:12.38.Mh; 24.85.+p; 25.75.-q}


\section{Introduction}
In ultrarelativistic heavy-ion collisions (HIC) at the Relativistic Heavy Ion Collider (RHIC) and the Large Hadron Collider (LHC) (commonly termed as mini big bangs), a new form of matter $-$ the quark-gluon plasma (QGP)\cite{QGP1,QGP2}, in which quarks, antiquarks, and gluons are deconfined, is created.  The large transverse momentum (high-$p_\perp$) particles are formed immediately upon the collision and therefore are affected by all stages of QGP evolution. This makes them excellent probes\cite{probe1, probe2} of this new state of matter, primarily through two main energy loss-based high-$p_\perp$ observables $-$ angular averaged ($R_{AA}$) and angular differential ($v_2$) nuclear modification factors.

Traditionally, rare high-$p_\perp$ probes ($p_\perp \gtrsim 5$ GeV), which present $\sim 0.1 \%$ of all particles produced in HIC, are used for studying the mechanisms of jet-medium interactions, while low-$p_\perp$ sector\cite{low1,low2,low3} ($p_\perp \lesssim 5$ GeV) is used to infer QGP features, such as e.g. initial stages before the QGP thermalization. However, up-to-date initial-stage properties are poorly known. Therefore, the need for an alternative approach to assessing the initial-stage features emerged. We here propose using high-$p_\perp$ probes as a complementary tool for this purpose, primarily since high-$p_\perp$ partons are good probes of QGP properties, where these properties depend on initial QGP stages.
Furthermore, the recently acquired extensive set of high-precision experimental data for the two aforementioned high-$p_\perp$ observables \cite{CMS,ATLAS,ALICE,CMS1,ATLAS1,ALICE1} facilitates our study. This issue is moreover intriguing, as results of current theoretical studies on this subject are mostly inconclusive\cite{IS1,IS2,IS3}.

A more rigorous study on this issue is required, that implies higher control over both the energy loss model and the analyzed temperature profiles. To accomplish this, we apply a full-fledged DREENA-B\cite{DREENAB} framework (B stands for one-dimensional (1D) Bjorken\cite{Bjorken} expansion), based on our state-of-the-art dynamical energy loss formalism\cite{RunA} that will be outlined in the next section. It also considers 1D Bjorken\cite{Bjorken} medium evolution, which is highly suitable for this study, as it allows the analytical introduction of different evolutions before thermalization, with the same evolution after thermalization, which facilitates the isolation of the effects of different initial stages. Additionally, we checked that the transition from 1D Bjorken to full 3+1D hydrodynamical evolution\cite{3D} does not significantly change our high-$p_\perp$ predictions, implying that for reliable high-$p_\perp$ predictions, an accurate energy loss model is more important than the medium evolution model. Therefore, DREENA-B\cite{DREENAB} provides an optimal framework for the initial-stages study, as it combines a state-of-the-art energy loss model with fully controlled temperature profiles. Note that, in this paper, we provide a part of the more detailed results obtained in Ref.~\citen{rad},  enriched with some complementary predictions.

\section{Numerical and Theoretical Framework}

For generating medium modified distribution of high-$p_{\perp}$ particles, irrespective of their flavor, we apply the generic pQCD  convolution formula~\cite{RunA,PLF}:
\begin{equation}
\frac{E_f d^3\sigma}{dp^3_f} = \frac{E_i d^3\sigma(Q)}{dp^3_i} \otimes P(E_i \rightarrow E_f)\ \otimes D(Q \rightarrow H_Q),
 \label{pQCD}
\end{equation}
where $i$ and $f$ stand for the initial parton ($Q$) and final hadron ($H_Q$), respectively. The initial parton momentum distribution $\frac{E_i d^3\sigma(Q)}{dp^3_i}$ is calculated in accordance with Ref.~\citen{ID}. The energy loss probability $P(E_i \rightarrow E_f)$ is based on our  dynamical energy loss formalism (see paragraph below) and incorporates multigluon\cite{MGF} and path-length fluctuations\cite{PLF,PLF1,DREENAC}. $D(Q \rightarrow H_Q)$ denotes fragmentation function, where for the light hadrons, D and B mesons de Florian-Sassot-Stratmann (DSS)\cite{DSS}, Braaten-Cheung-Fleming-Yuan (BCFY)\cite{BCFY} and Kartvelishvili-Likhoded-Petrov (KLP)\cite{KLP} fragmentation functions are used, respectively.

As a crucial ingredient of the calculations, we employ our state-of-the-art dynamical energy loss formalism\cite{DRad,DRad1,DColl}, which includes: 1) Dynamical QCD medium of a finite temperature and a finite size, so that the energy exchange with the medium constituents is taken into account as opposed to static scattering centers case. It also considers that the medium created in ultrarelativistic heavy-ion collisions has a finite size, and that initial partons are created inside the medium. 2) The calculations are based on the finite temperature field theory and generalized hard-thermal-loop approach\cite{Kapusta}, generically regulating infrared divergences. 3) Both collisional\cite{DColl} and radiative\cite{DRad,DRad1} energy loss mechanisms are included and performed within the same theoretical framework, so that no interference or overlapping occurs between them. 4) The formalism is generalized to the case of finite magnetic mass\cite{MM} and running coupling\cite{RunA}. Chromomagnetic ($\mu_M$) to chromoelectric mass ($\mu_E$) ratio is estimated to be in a range $0.4 - 0.6$ by different non-perturbative approaches\cite{xb,xb1}. Hence, in this paper we assume $\mu_M / \mu_E = 0.5$. Our most recent advancement within formalism is the relaxation of the widely used soft-gluon approximation\cite{bsga}. In Ref.~\citen{ELEf}, we demonstrated that all the above ingredients are necessary to accurately reproduce high-$p_\perp$ suppression data.

The full-fledged analytical expressions for single gluon radiation spectrum and collisional energy loss per unit length in an expanding medium are given by Eqs. (6) and (2) from Ref.~\citen{DREENAB}, respectively. Thereby, the standard values for heavy and light quark masses are considered ($M_c = 1.2$ GeV, $M_b = 4.75$ GeV, while for light quarks thermal masses are assumed).

Further, we assume that the medium expansion model is given by the ideal hydrodynamical 1D Bjorken expansion\cite{Bjorken}, {\it i.e.},  $T(\tau) \sim \sqrt[3]{(\tau_0 / \tau)}$ (where $\tau$ is a proper time), with thermalization time set at $\tau_0 = 0.6$ fm/c\cite{tau0}. The detailed determination of initial QGP temperature $T_0$ for the considered centrality range is provided in Ref.~\citen{DREENAB}. For brevity, here we focus on  $30-40\%$ centrality region in $\sqrt{s_{NN}}=5.02$ TeV $Pb+Pb$ collisions  at the LHC, which corresponds to $T_0=391$ MeV\cite{DREENAB}, although we checked that the same conclusions apply regardless of the considered centrality bin.  The QGP transition temperature is assumed to be $T_C \approx 160$\cite{Temp0}.

Finally, we provide the expressions for two main high-$p_\perp$ observables. The angular averaged nuclear modification factor
 $R_{AA}$ is defined as the ratio of the quenched $p_{\perp}$-spectrum in $A + A$ collisions with respect to $p + p$ collisions, normalized
by the number of binary collisions $N_{bin}$:
\begin{equation}
R_{AA}(p_T)=\dfrac{dN_{AA}/dp_T}{N_{\mathrm{bin}} dN_{pp}/dp_T}.
\label{RAA}\
\end{equation}
However, an alternative form\cite{DREENAC}
\begin{equation}~\label{RAA0}
R_{AA}\approx \frac{R^{in}_{AA} + R^{out}_{AA}}{2},
\end{equation}
is also used for providing more intuitive insight in the underlying mechanisms. Here\cite{IS1} $R^{in}_{AA}=\dfrac{dN_{AA}/dp_T d \phi \mid_{\phi=0}}{N_{\mathrm{bin}} dN_{pp}/dp_T d \phi \mid_{\phi=0}}$ ($R^{out}_{AA}=\dfrac{dN_{AA}/dp_T d \phi \mid_{\phi=\pi /2}}{N_{\mathrm{bin}} dN_{pp}/dp_T d \phi \mid_{\phi=\pi /2}}$) stands for in-(out-of-)plane nuclear modification factor.
The  high-$p_{\perp}$ elliptic flow is given by the expression\cite{DREENAC, v2f,Betz}:
\begin{equation}
v_{2} \approx \frac{1}{2} \frac{R^{in}_{AA} -R^{out}_{AA}}{R^{in}_{AA} + R^{out}_{AA}}.
\label{v20}
\end{equation}

It is worth noting that experimental approach to $v_2$  is different from Eq.~(\ref{v20}). However, to our knowledge, and as already discussed in Ref.~\citen{rad}, that approach could lead to different elliptic flow predictions if event-by-event fluctuations are taken into account, which is out of the scope of this study. 

\section{Reliability of the Framework}

The reliability of DREENA-B framework, outlined in the previous section, is tested against experimentally available data at the LHC in Ref.~\citen{DREENAB}. Note that in generating all predictions we used no fitting parameters, {\it i.e.}, the parameters take their standard literature values.  We obtained a very good agreement between our predictions and the existing data for: {\it i)} Both high-$p_\perp$ $R_{AA}$ and $v_2$, so that long-standing $v_2$ puzzle\cite{puzzle} (inability of various models to jointly explain high-$p_\perp$ $R_{AA}$ and $v_2$ data, with tendency to underestimate $v_2$ compared to the experimental data) is naturally solved within our framework; {\it ii)} Diverse colliding systems, such as Pb + Pb at $\sqrt{s_{NN}}= 2.76$ TeV and 5.02 TeV, and Xe + Xe at  $\sqrt{s_{NN}}=5.44$ TeV; {\it iii)} Both light and heavy flavor particles, that is, $h^\pm, D, B$ mesons, and {\it iv)} All available centrality ranges.

\section{Results and Discussion}

In this section, first, we define the four commonly considered temperature profiles\cite{IS1}, which differ only at early times. Next, we assess their effects on our full-fledged predictions for high-$p_\perp$ angular averaged and angular differential nuclear modification factors. Finally, we revise the soundness of the commonly applied multiple fitting procedure. For each result, we provide an intuitive explanation based on $R_{AA}$ asymptotic scaling behavior.  For more details, we refer the reader to Ref.~\citen{rad}.

\begin{figure}[th]
\centerline{\includegraphics[width=0.8\textwidth]{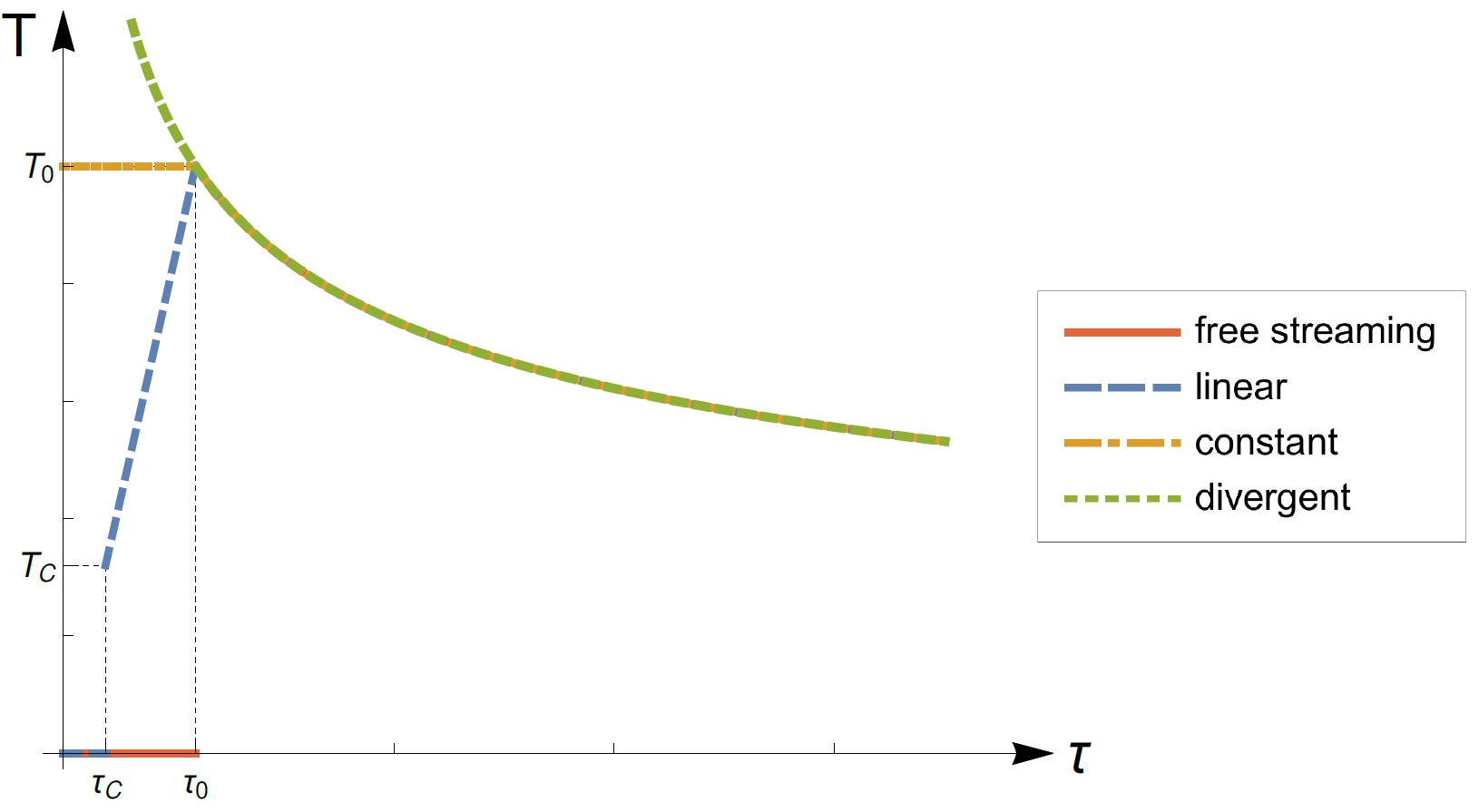}}
\caption{Four simplified temperature profiles, with the same 1D Bjorken\cite{Bjorken} temperature evolution after thermalization ($\tau \geq \tau_0$), and whose differences before thermalization mimics different evolutions at initial stage ($\tau < \tau_0$). These diverse initial-stage cases are: the {\it free-streaming} (full red curve),  the {\it linear} (dashed blue curve), the {\it constant} (dot-dashed orange curve) and {\it divergent} case (dotted green curve), as denoted in the legend.  Figure adapted from Ref.~\citen{rad}.}
\end{figure}

\subsection{Effect of different initial stages on high-$p_\perp$ $R_{AA}$ and $v_2$}

Now that framework is set and tested, for the study covered by this paper, next we concentrate on four commonly assumed temperature profiles\cite{IS1} that consider the same 1D Bjorken\cite{Bjorken} temperature profile after, but differ before the thermalization ($\tau < \tau_0$). More particularly, in Fig. 1, we distinguish:
\begin{itemlist}
\item the {\it free-streaming} case (full red curve), which corresponds to omitting the energy loss before the QGP thermalization;
\item  the {\it linear} case (dashed blue curve), which corresponds to linearly increasing $T$ with proper time from transition temperature ($T_C = 160$ MeV\cite{Temp0}, $\tau_C = 0.25$ fm) to the initial temperature $T_0$ of equilibrated plasma, otherwise $T=0$;
\item the {\it constant} case (dot-dashed orange curve), with $T$ equal to the initial temperature $T_0$; and
\item the {\it divergent} case (dotted green curve), corresponding to 1D Bjorken evolution from the beginning $\tau = 0$.
\end{itemlist}

First, we assess to what extent high-$p_\perp$  $R_{AA}$ is affected by the presumed initial stages depicted in Fig. 1. From the left column of Fig. 2, we infer that high-$p_\perp$  $R_{AA}$ is sensitive to the initial stages. Particularly, we see that for both light and heavy flavor particles, suppression is the lowest in the {\it free-streaming} case, while progressively increasing toward the {\it divergent} case, which is expected due to an increase in energy loss. Unfortunately, the discrepancies between these curves are not very distinguishing, and within the current error bars, one is unable to differentiate between these different scenarios.

\begin{figure}[th]
\centerline{\includegraphics[width=0.9\textwidth]{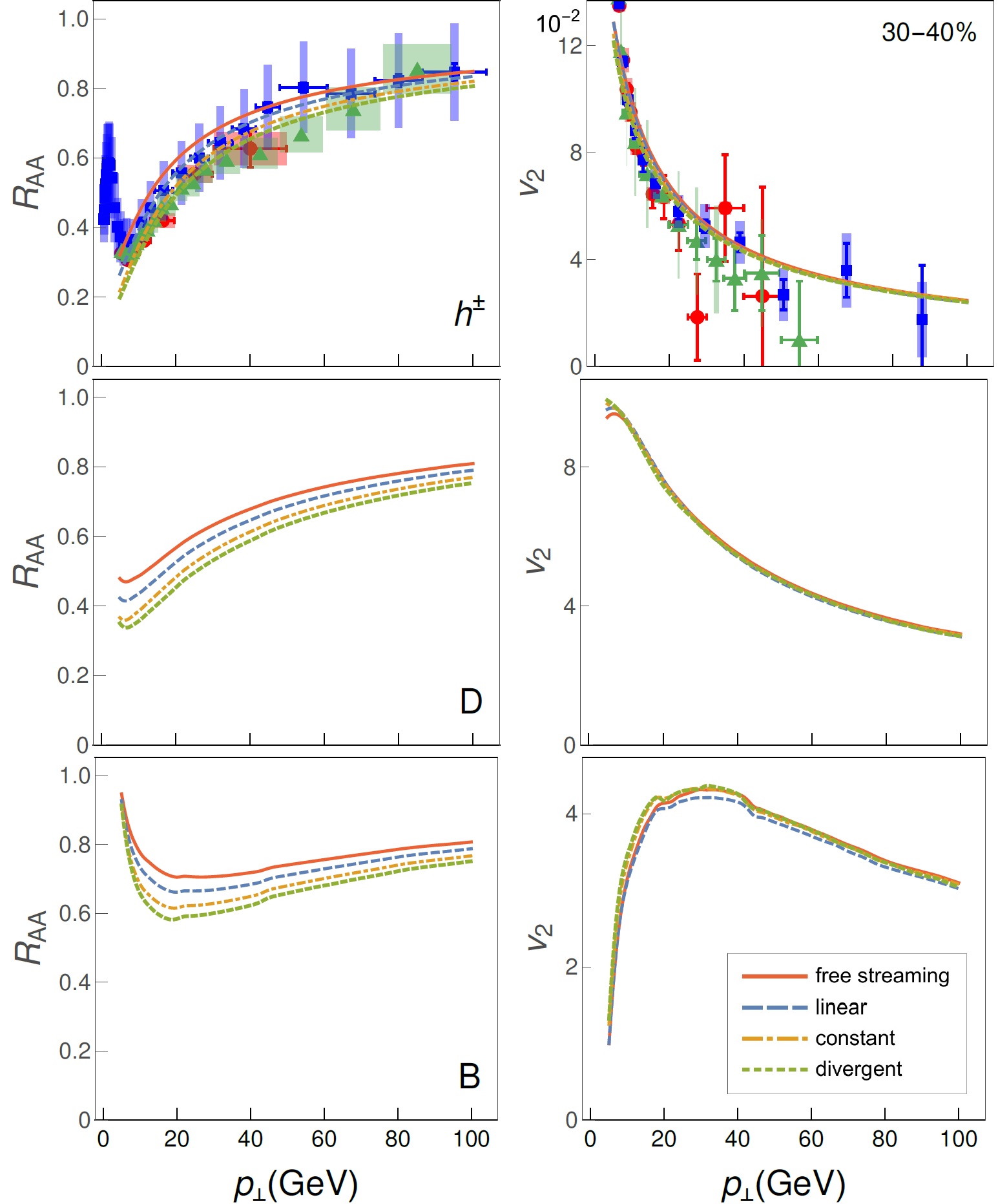}}
\caption{Sensitivity of high-$p_\perp$ observables  to different initial stages presented in Fig. 1. The left (right) column  corresponds to high-$p_\perp$ $R_{AA}$ ($v_2$) {\it vs} $p_\perp$.  Charged hadron, D meson and B meson predictions are presented in upper, middle and lower row, respectively. Charged hadron $R_{AA}$ predictions are compared with 5.02 TeV Pb + Pb  CMS\cite{CMS} (blue squares), ATLAS\cite{ATLAS} (green triangles) and ALICE\cite{ALICE} (red circles) $h^\pm$ $R_{AA}$ data in the upper left plot, while its $v_2$ predictions are compared with the corresponding 5.02 TeV Pb + Pb  CMS\cite{CMS1} (blue squares), ATLAS\cite{ATLAS1} (green triangles) and ALICE\cite{ALICE1} (red circles) $h^\pm$ data in the upper right plot. In each plot, full red curve corresponds to the {\it free-streaming} case, dashed blue curve to the {\it  linear} case, dot-dashed orange curve to the {\it constant} case, and  dotted green curve to the {\it divergent} case, as indicated in legend. The results are presented for
the centrality range $30-40 \%$, and $\mu_M / \mu_E=0.5$. Figure adapted from Ref.~\citen{rad}.}
\end{figure}

Next, we investigate the sensitivity of high-$p_\perp$ elliptic flow to the initial stages. Unexpectedly, from the right column of Fig. 2, we observe that $v_2$ is insensitive to the presumed initial stages for all types of particles, contrary to the conclusion derived in Ref.~\citen{IS3}. Therefore, from our study, it follows that $v_2$ cannot differentiate between different initial-stage scenarios.

\begin{figure}[th]
\centerline{\includegraphics[width=0.9\textwidth]{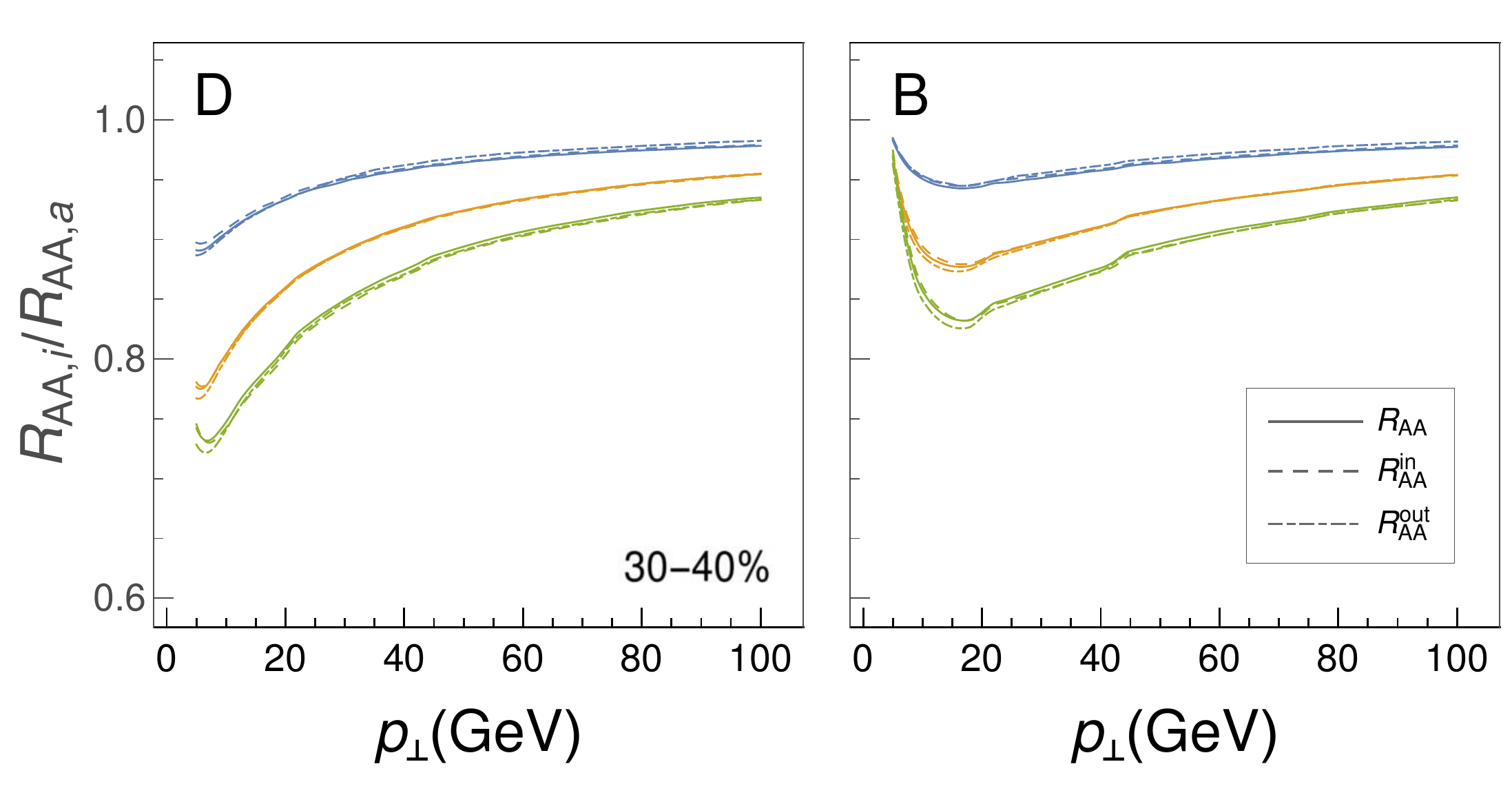}}
\caption{$R^{in}_{AA}$ (dashed curves), $R^{out}_{AA}$ (dot-dashed curves) and $R_{AA}$ (full curves) {\it vs} $p_\perp$ in {\it linear} (blue set of curves), {\it constant} (orange set of curves) and {\it divergent} case (green set of curves) relative to the {\it free-streaming} case. The left (right) plot corresponds to D (B) mesons. The parameters are the same as in Fig. 2.}
\end{figure}

To quantitatively explain the obtained results, we plot heavy flavor momentum dependence of proportionality functions, which are defined in the following manner:
\begin{equation}
\gamma^{in}_{i}= \frac{R^{in}_{AA,i}}{R^{in}_{AA,fs}},\quad
\gamma^{out}_{i} = \frac{R^{out}_{AA,i}}{R^{out}_{AA,fs}}, \quad
\gamma_{i} = \frac{R_{AA,i}}{R_{AA,fs}},
\label{in}
\end{equation}
where $i\in \{lin, const, div \}$. The results and conclusions for charged hadrons are the same and are shown in Ref.~\citen{rad}. Thus, in Fig. 3, we distinguish three sets of curves (corresponding to {\it linear, divergent}, and  {\it constant} cases relative to {\it free-streaming} case), each of which contains corresponding three proportionality functions. The most important observation from Fig. 3  is that within the same set of curves  the proportionality functions are practically identical for the relations involving $R^{in}_{AA}$, $R^{out}_{AA}$ and $R_{AA}$, that is
 \begin{equation}
\gamma^{in}_{i} \approx \gamma^{out}_{i} \approx \gamma_{i}.
\label{inout}
\end{equation}
It is worth noting that $\gamma_{i} < 1$, and that for $i \neq j \rightarrow \gamma_{i}(p_{\perp}) \neq \gamma_{j}(p_{\perp})$.
If we recall that high-$p_\perp$ $R_{AA}$ and $v_2$ are  given by Eqs.~(\ref{RAA0}) and~(\ref{v20}), it is straightforward to show that only $R_{AA}$, and not $v_2$ is affected. More specifically, for any $i$ we obtain:

\begin{equation}
R_{AA,i}  \approx \frac{\gamma_{i} (R^{in}_{AA,fs} + R^{out}_{AA,fs})}{2} =\gamma_i R_{AA,fs},
\label{RAAp}
\end{equation}

\begin{figure}[th]
\centerline{\includegraphics[width=0.9\textwidth]{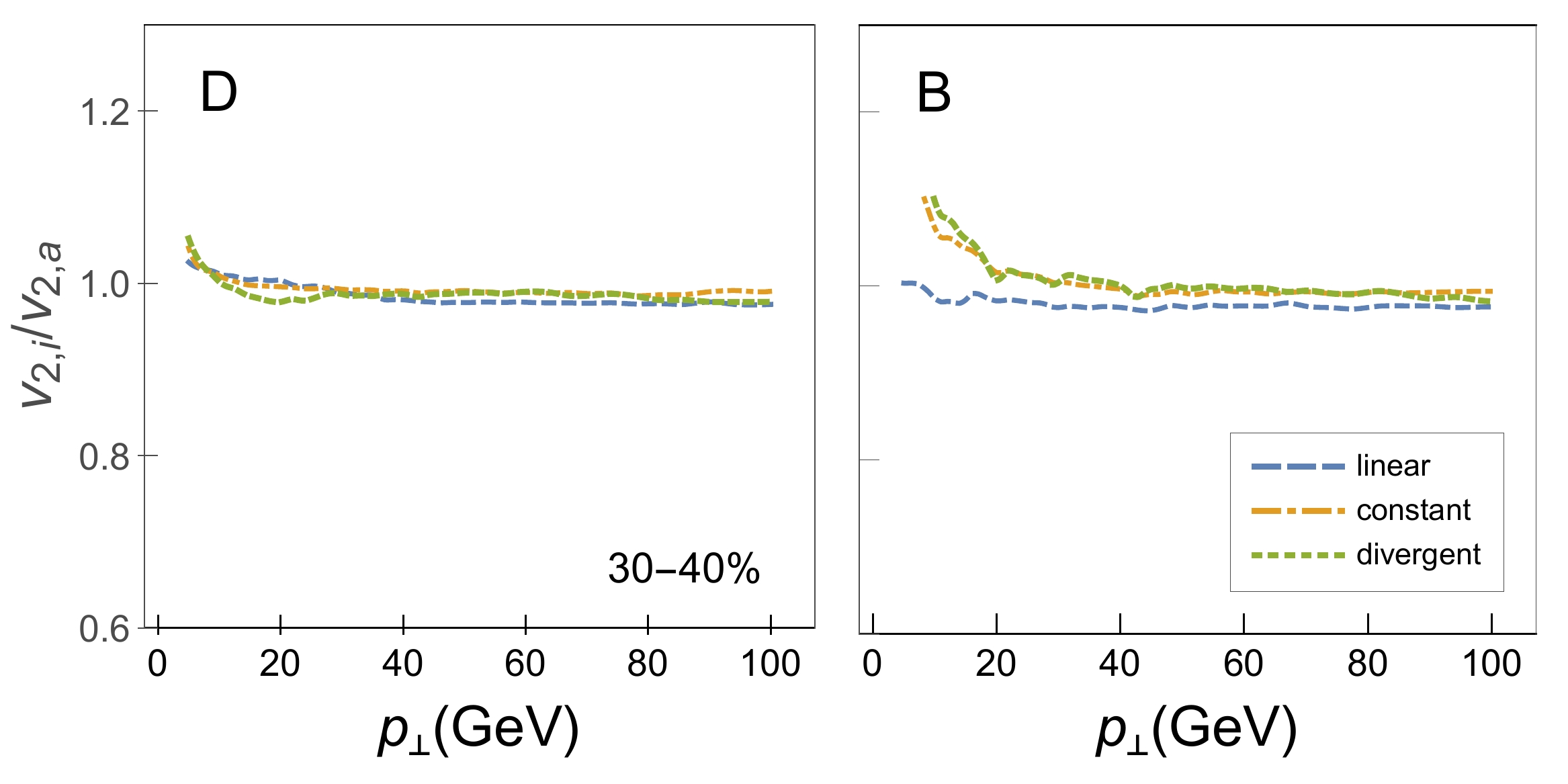}}
\caption{$v_2$ in {\it linear} (dashed blue curves), {\it constant} (dot-dashed orange curves) and  {\it divergent} case (dotted green curves) {\it vs} $p_\perp$ relative to the {\it free-streaming} case. The left (right) plot corresponds to D (B) mesons. The parameters are the same as in Fig. 2.}
\end{figure}

\begin{equation}
v_{2,i}  \approx \frac{1}{2} \frac{\gamma_{i} (R^{in}_{AA,fs} - R^{out}_{AA,fs})}{\gamma_{i} (R^{in}_{AA,fs} + R^{out}_{AA,fs})} =v_{2,fs},
\label{v2p}
\end{equation}
 as observed in Fig. 2.

As an additional test of $v_2$ equivalence for different initial stages, in Fig. 4, we present the ratio of elliptic flow in {\it linear, constant} and {\it divergent} cases relative to the {\it free-streaming} case. From Fig. 4, which is a counterpart of Fig. 3, we observe that these ratios are consistent with unity for both D and B mesons. The result is the same for charged particles and omitted for consistency. Note that our predictions are valid for $p_{\perp} \gtrsim$ 10 GeV. This, furthermore, confirms the conclusions obtained from Fig. 2 (right column) and Fig. 3, as well as the validity of our quantitative analysis (given by Eqs. (7) and (8)).

 Additionally, $R_{AA}$ sensitivity to the initial stages is in a qualitative agreement with Refs.~\citen{DREENAB,Renk, Molnar}, where it was shown that high-$p_\perp$ $R_{AA}$ is only sensitive to the averaged properties of the evolving medium,  such as average $T$ ($\overline{T}$), {\it i.e.}, the analytical estimate reads:
 \begin{equation}
R_{AA}  \sim \frac{\Delta E}{E} \sim \overline{T}.
\label{RAAapp}
\end{equation}
 The fact that $\overline{T}$s are different for all four initial-stage cases (see Fig. 1) results in observed $R_{AA}$ differences.

\subsection{Revision of commonly used multiple fitting procedure}
Finally, in this subsection, we test an approach commonly used in Refs.~\citen{IS3, Noronha, Betz, Betz1, Prado, Cao}, in which the energy loss is fitted for the initial-stage cases (see Fig. 1), via the change of multiplicative fitting factor in the energy loss to reproduce the high-$p_\perp$ $R_{AA}$ experimental data. More specifically, in our full-fledged calculations we introduce an additional multiplicative fitting factor $C^{fit}_i$, which is estimated for each initial-state case as the best fit to the {\it free-streaming} $R_{AA}$, since free-streaming is commonly considered scenario in both low- and high-$p_\perp$ sector\footnote{The estimated\cite{rad} values of $C^{fit}_i$ are: 1, 0.87, 0.74 and 0.67 in {\it free streaming, linear, constant} and {\it divergent} cases, respectively.}. We observe\cite{rad} a decreasing trend in multiplicative fitting factors from the {\it free steaming} toward the {\it divergent} case, as expected, to annul the higher energy losses in corresponding cases compared to the {\it free-streaming} one.

Thus obtained (fitted) $R_{AA}$s are presented in the left plot of Fig. 4, and are practically overlapping, as expected. However, the right plot of Fig. 4 shows that through this fitting procedure  high-$p_\perp$ $v_2$ is significantly affected, that is, the highest value is in the {\it free-streaming} case, while the lowest is in the {\it divergent} case. This observation could evoke a naive interpretation that initial stages, that is, the only region in which $T$ profiles differ, are responsible for these discrepancies. However, that would be inconsistent with our results presented in the previous subsection, as well as with intuitive anticipation that the introduction of the energy loss at the initial stage should affect $R_{AA}$.

\begin{figure}[th]
\centerline{\includegraphics[width=\textwidth]{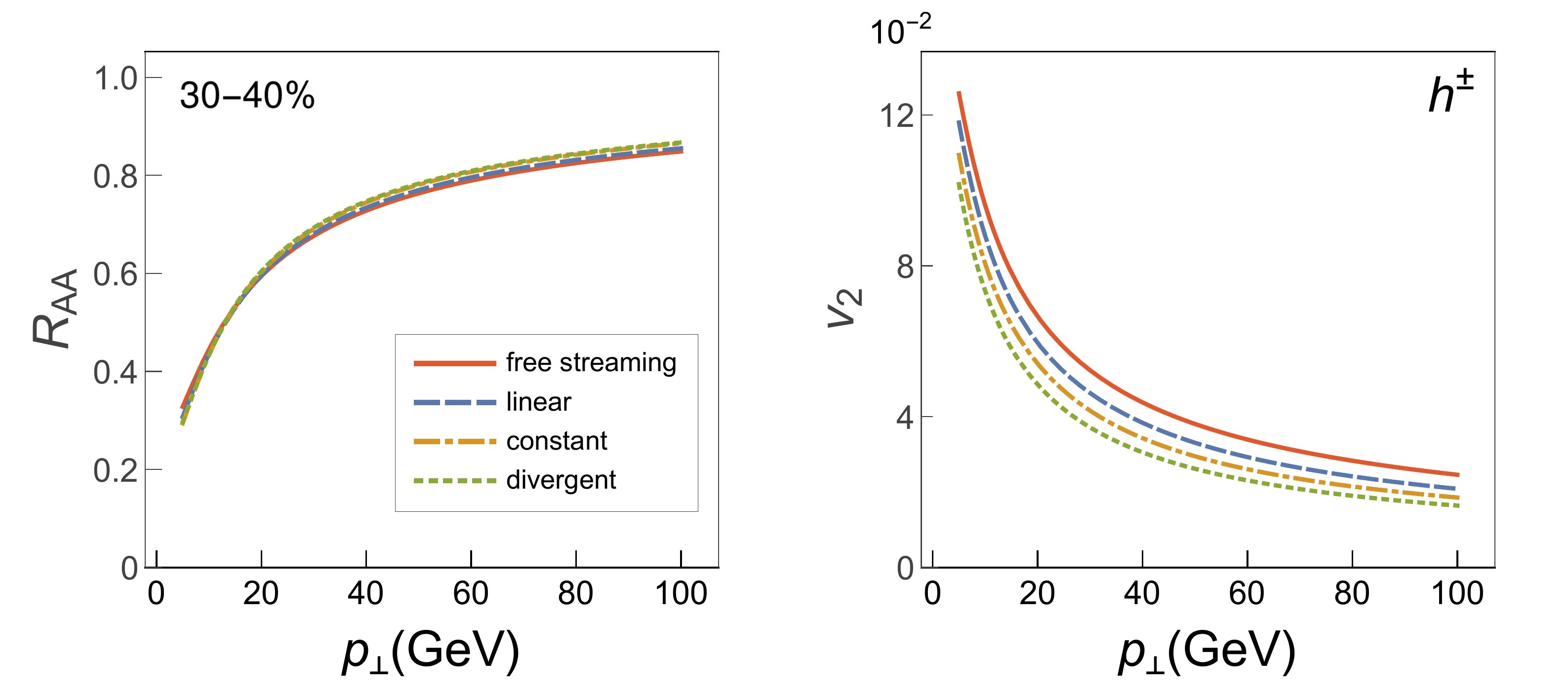}}
\caption{Sensitivity of charged hadron high-$p_\perp$ $R_{AA}$ (left plot) and $v_2$ (right plot) to
different initial-stage cases from Fig. 1, when a multiplicative factor is included in energy loss to reproduce the {\it free-streaming} $R_{AA}$. In each plot, full red curve corresponds to the {\it free-streaming} case, dashed blue curve to the {\it fitted  linear} case, dot-dashed orange curve to the {\it fitted constant} case, and  dotted green curve to the {\it fitted divergent} case, as indicated in legend. The  parameters are the same as in Fig. 2. Figure adapted from Ref.~\citen{rad}.}
\end{figure}
To provide a quantitative explanation of the obtained results in Fig. 4, we apply asymptotic scaling behavior of $R_{AA}$~\cite{DREENAB,DREENAC}, which mimics our complex suppression procedure for very high-$p_\perp$ jets and at higher centralities:
\begin{equation}
R_{AA} = 1-\xi \overline{T}^m \overline{L}^n,
\label{est}
\end{equation}
where $\overline{L}$ denotes the average path length traversed by the jet. The corresponding $\overline{T}$ and $\overline{L}$ proportionality factors $-$ $m\approx 1.2$ and $n \approx 1.4$ are estimated in Refs.~\citen{m} and ~\citen{n}, respectively. $\xi$ stands for a proportionality factor, which depends on jet's $p_\perp$ and flavor.

  By introducing multiplicative fitting factor in energy loss (see Eq.~(\ref{RAAapp})), and making use of Eq.~(\ref{est}), the fitted $R_{AA}$s in high-$p_\perp$ limit now read:
 \begin{equation}
R^{fit}_{AA,i} \approx 1-C_{i} \xi \overline{T}^m_{i} \overline{L}^n_{i} \approx 1-C_{i} (1- R_{AA, i}),
\label{R31}
\end{equation}
where $i= lin, const, div$ and $C_i$s are high-$p_\perp$ limits of corresponding $C^{fit}_i$s. Note that Eqs.~(\ref{est}) and (\ref{R31}) (in their original form) are applicable to  $R^{in}_{AA}$ and $R^{out}_{AA}$ as well (the same multiplicative fitting factor is naturally  applied in all three cases).  
In order for $R^{fit}_{AA,i}$ to reproduce the free streaming $R_{AA}$, {\it i.e.},
\begin{equation}
R^{fit}_{AA,i} = R_{AA,fs},
\label{R32}
\end{equation}
 it is straightforward to obtain:
\begin{align}~\label{v3}
v^{fit}_{2,i} {} &  \approx  \frac{1}{2} \frac{C_{i} (R^{in}_{AA,i} - R^{out}_{AA,i})}{2 R_{AA,fs}} = \frac{1}{2} \frac{C_{i} \gamma_{i}(R^{in}_{AA,fs} - R^{out}_{AA,fs})}{R^{in}_{AA,fs} + R^{out}_{AA,fs}} \nonumber \\
&  = C_{i} \gamma_{i} v_{2,fs},
\end{align}
where along with Eq.~(\ref{R32}), we applied Eqs.~(\ref{RAA0})-(\ref{inout}), and Eqs.~(\ref{est}), (\ref{R31}), together with their out- and in-plane counterparts.

From Eq.~(\ref{v3}) it follows that the reasons behind $v_2$ decrease in {\it linear, constant} and {\it divergent} cases compared to the {\it free streaming} one are the multiplicative fitting factor $C_i$ and proportionality function $\gamma_i$, both of which are smaller than 1. However, note that $\gamma_i$ approaches to 1 at high $p_\perp$ (we refer the reader to Fig. 3), so that the diminishing of $v_2$ compared to the {\it free streaming} case is predominantly a consequence of a decrease in the imposed fitting factor and not the initial stages as obtained in Ref.~\citen{IS3}. We thus infer that the common procedure in which the energy loss fitting factor is repeatedly adjusted for each initial stage may lead to misconceptions about the underlying physical mechanisms.

\section{Conclusions and Outlook}

We here addressed whether, and to what extent, we can use high-$p_\perp$ observables to explore the initial stages before QGP thermalization. To this end, we studied how four different commonly considered initial stage scenarios, which have the same temperature profile after, but differ in the temperature profiles before thermalization, affect high-$p_\perp$ $R_{AA}$ and $v_2$ predictions, stemming from our DREENA-B\cite{DREENAB} framework combined with 1D Bjorken expansion\cite{Bjorken}. We surprisingly obtained that high-$p_\perp$ $v_2$ is insensitive to the presumed initial stages, as opposed to high-$p_\perp$ $R_{AA}$. However, within the current error bars, $R_{AA}$ sensitivity does not allow differentiation between different initial stage cases. Moreover, we inferred that the previously reported sensitivity of high-$p_\perp$ $v_2$ to initial stages is mostly an artifact of the fitting procedure. Consequently, a common procedure, where free parameters in energy loss are separately fitted for each initial stage may obscure the understanding of the underlying physical mechanisms. In general, our results imply that the simultaneous study of high-$p_\perp$ $R_{AA}$ and $v_2$, with restrained temperature profiles (isolating the differences in the initial states) and unchanged energy loss parametrization throughout the study, is needed to set reliable constraints on the initial stages in the future.

\section*{Acknowledgements}

This work is supported by the European Research Council, grant ERC-2016-COG: 725741, and by the Ministry of Science and Technological Development of the Republic of Serbia, under project No. ON171004 and No.  ON173052.

\end{document}